# Formation and stability of icosahedral phase in $Al_{65}Ga_5Pd_{17}Mn_{13}$ alloy

**T.P. Yadav, M.A.Shaz, R.S.Tiwari and O.N.Srivastava**[*]

Department of Physics, Banaras Hindu University, Varanasi-221005, India

**Abstract**

In this work, we present the formation and characterization of a quaternary (pseudo ternary) icosahedral quasicrystal in $Al_{65}Ga_5Pd_{17}Mn_{13}$ alloy. The X-ray diffraction and transmission electron microscopy confirmed the formation of icosahedral, B2 type and ξ' crystalline (orthorhombic structure with unit cell a = 23.5Å, b= 16.6Å and c = 12.4 Å) phases in as cast alloy. The icosahedral phase gets formed after annealing at 800 °C for 60 hours. The formation of icosahedral phase in Al-Ga-Pd-Mn quaternary alloy by present technique has been studied for the first time. The Energy dispersive X-ray analysis investigations suggest the presence of Ga (~ 5 at %) in the alloy. It is interesting to note that pseudo twelve fold pattern in the as cast alloy has been observed. Icosahedral Al–Ga–Pd–Mn provides a new opportunity to investigate the various characteristics including surface characteristics. Attempts will be made to discuss the micromechanisms for the formation of quasicrystalline phase in Al-Ga-Pd-Mn alloys.

Corresponding authors: hepons@yahoo.com

# 1. Introduction

Quasicrystals are very interesting and comparatively new phase of solid discovered by Shechtman et al. in 1984 [1], which are intensively studied due to their unusual properties and promising perspectives for possible applications. Quasicrystalline materials have been discovered over 25 years ago, yet much debate remains on the origins of their desirable materials properties. These desirable properties reportedly include low friction, high hardness, high wear and corrosion resistance, and low surface energy [2–9]. Quasicrystals were also found to have low thermal conductivity and moderate electrical conductivity, making them suitable for use in thermoelectric devices [10-12]. The most promising characteristic of quasicrystals is their inherently low thermal conductivity (1-3 W/m×K), it is due to essentially infinite unit cell size and scattering on large heavy ions [13]. At room temperature, the best sample of the AlPdMn quasicrystalline system shows reasonable thermal conductivity values (500 – 800 (W×cm)-1) and thermopower values (80 mV/K) [13]. In Al- Pd- Mn alloy system, three types of quasicrystals, namely a stable decagonal phase with 1.2 nm periodicity, a stable icosahedral phase, and a metastable decagonal phase with 1.6 nm periodicity, successively appear with increasing Pd content in composition $Al_3(Mn_{1-x}Pd_x)$ (Tsai *et al.* 1989, Beeli *et al.* 1991, Hiraga *et al.*1991) [14-16]. Experimental results show that the i-phase in Al-Pd-Mn alloy belongs to the face-centered icosahedral (FCI), similar to that in Al-Cu-Fe alloy and it remains stable up to the melting point. A thermodynamically stable approximant phases have been found in the compositional vicinity of quasicrystals in a number of quasicrystal-forming alloy systems. In the Al-Pd-Mn alloy system, besides icosahedral and decagonal phases, some orthorhombic (ξ') phases have been reported, which were shown to be quasicrystal approximants [17]. The stability range of the ξ phase, which has a structure closely related to that of the binary $Al_3Pd$ phase (Matsuo and Hiraga 1994) [18], is located in the low-Mn region. Considering the local structure of the material, ξ' (Al-Pd-Mn) is recognized to be an approximant of the well-known icosahedral phase $Al_{70.5}$ $Pd_{21}Mn_{8.5}$. In particular, the icosahedral phase in Al-Pd-Mn has Mackay-type clusters as elementary structural building blocks (Boudard *et al.* 1996) [19]. The space group of the ξ' is P*nma* and the unit cell containing 316 atoms has lattice parameters *a* =23.54 Å, *b* =16.56Å and *c* =12.34 Å.

The formation of quasicrystalline phase in Al-Ga-Pd-Mn system is difficult due to the low melting temperature of Ga (29.76°C). A quaternary icosahedral *i*-Al-Ga-Pd-Mn quasicrystal (QC) with composition $Al_{67}Ga_4Pd_{21}Mn_8$ was discovered in 1999 by Fisher and co-workers,[20] following attempts to grow ternary *i*-Al-Pd-Mn from a fourth element flux in this case Ga. Selected area diffraction patterns (SADPs) suggested that this new QC is isostructural with *i*-Al-Pd-Mn, where 4 at % Ga gets incorporated into the lattice. The quasilattice dimension of *i*-Al-Ga-Pd-Mn was found approximately 1.5% larger than that of the ternary compound. Apart from the original work, [21-22] the investigations of the structure and physical properties of the *i*-Al-Ga-Pd Mn QC are still scarce, therefore some basic issues remain open. It is not clear whether the Ga atoms occupy specific sites in the lattice or the Ga-Al substitution can be at random, i.e., whether this is a true quaternary or a pseudoquaternary QC. In addition, the reported electrical resistivity measurements suggest that as little as 4% Ga in the *i*-Al- Pd-Mn structure increases the resistivity by about 50%. The formation of QC phase in Al-Ga-Pd-Mn alloys system by simple melting and annealing processes has not been reported.

In the present paper, the formation of icosahedral quasicrystalline phase in $Al_{65}Ga_5Pd_{17}Mn_{13}$ alloy by melting and annealing prosses will be reported with an aim to understand the microstructural evolution and phase stability during annealing.

## 2. Experimental details

The alloy with nominal composition of $Al_{65}Ga_5Pd_{17}Mn_{13}$ were prepared in an argon atmosphere by melting high purity Al (99.96%), Ga (99.99%), Pd (99.96%) and Mn (99.99%) in a silica crucible using RF induction furnace The individual elements were taken in correct stoichiometric proportions and pressed into a cylindrical pellet of 1.5 cm diameter, 1 cm thickness by applying a pressure of ~ $2.76 \times 10^4$ N/m$^2$. The pellet (5g by weight) was then placed in a silica tube surrounded by an outer Pyrex glass jacket. Under continuous flow of argon gas into the silica tube, the pellet was melted using radio frequency induction furnace (18 kW). During melting process, water is circulated in the outer jacket around the silica tube to reduce the contamination of the alloy with silica tube and also for good cooling. The melting atmosphere was purified by previously melting Ti buttons. The as-cast ingot was melted several times to ensure homogeneity. The homogenized alloy was then subjected to annealing at 700 & 800 °C. Before annealing  the  as cast ingots were sealed in a silica tube, which was flushed two to three times with high purity argon gas and then evacuated to a pressure of ~ $1.32 \times 10^{-6}$ atm, It was then placed in a furnace and annealed at 600°C ~ 900$^\circ$C (±10$^\circ$C) for time spans ranging from 10 to 80 h. It was found that annealing at 800 °C for 60 h produced the optimum material. Therefore, this time period was maintained for all further annealing runs. The as cast and heat-treated alloy was characterized by powder X-ray diffraction (XRD) using a Philips 1710 X-ray diffractometer with CuK$\alpha$ radiation ($\lambda$=1.54026Å). The surface microstructure of as cast and annealed alloy was characterized through scanning electron microscopy (SEM) (Philips: XL 20). The phase transformation temperatures were made with Linesis L72 differential thermal analyzer (DTA) in vacuum at heating rate of 10°K/S. The transmission electron microscopy (TEM) using Philips EM CM-12 has been used for microstructural and structural characterizations with an operating voltage at 100 kV. An energy-dispersive X-ray link with HRTEM Tecnai 20 G$^2$ system was employed for the compositional analysis.

# 3. Results and Discussion

In order to explore the surface morphology of as cast and annealed $Al_{65}Ga_5Pd_{17}Mn_{13}$ alloy, scanning electron microscopy (SEM) has been carried out. Fig.1(a) shows SEM image of the as cast $Al_{65}Ga_5Pd_{17}Mn_{13}$ alloy. It can be seen that there are two types of contrast with spherical microstructure (marked as A, & B). A consists of long and straight rod like growth with diameter ranging from 30 to 40 of micrometers. However, the spherical type of growth on the surface of the rod with lengths from 6 to 10 of micrometers can be seen in fig 1(a). The morphology of fractured surface was obviously different from spherical microstructure. After 60 h of annealing at 800°C of as cast alloy a very smooth dodecahedran like microstructure has been observed (shown in fig. 1(b)). The dodecahedral microstructure of various quasicrystlline materials have been described extensively in the literature [23]. Therefore, it can be concluded that these small grains are icosahedral quasicrystalline phase.   In addition to dodecahedral grain there are some grain having polygonal/petal morphology as well.

Fig.2 shows a differential thermal analysis (DTA) scan of as cast $Al_{65}Ga_5Pd_{17}Mn_{13}$ alloy. An endothermic peak at approximately 760°C is clearly evident. As we will discuss later, this endothermic peak may be related to the transformation of crystalline phases, namely ξ` orthorhombic phase and  B2 phase to icosahedral phase. This indicates that i phase is more stable than the crystalline phase above 760 $^0$C.

Figure 3(a) shows the XRD pattern of as cast $Al_{65}Ga_5Pd_{17}Mn_{13}$ alloy indicating the formation of icosahedral ( I) phase, ξ` orthorhombic structure with unit cell a = 23.5Å, b= 16.6Å and c = 12.4 Å  and AlPd type  B2 phase.

The B2 phase is a approximant phase of quasicrystal, having a crystalline CsCl-type structure and nominal composition $Al_{48}Pd_{42}Mn_{10}$ along with those of an Al–Pd–Mn icosahedral quasicrystal [24]. The gradual formation of I phase after annealing for 60 h has been shown in fig 3(b-c) . There are significant changes in XRD pattern after annealing of as cast alloy up to 60 h, at both the temperatures 700&800°C which can be inferred from variation in peak intensity and broadening. In the case of 700°C annealing the B2 phase continued to be present with i phase and ξ` phase. After 60 h of annealing at 800°C, IQC phase has been observed (fig.3c) as dominant one with very small amount of ξ' phase . It was found that annealing of the as cost alloy at 900°C resulted into the decomposition of I phase into $Al_3Pd_2$ and icosahedral phase. The corresponding XRD pattern is not given for the sake of brevity.

The compositional analysis of as cast $Al_{65}Ga_5Pd_{17}Mn_{13}$ was carried out by energy depressive X-ray analysis (EDX) attached to the transmission electron microscope Tecnai G[20] ( sown in Fig.4).The EDX analysis shows the presence of all the elements in as cast alloy (Al= 65.2 at%, Ga=4.6 at%, Pd=18.2 at% and Mn= 12 at%) and it is very close to stoichiometric proportions.  It also shows the very small presence of oxygen approximated to be around 0.14 at % in the sample. However, after annealing experiment the oxygen contamination increases up to 0.2 at % indicating the small oxygen pick up during annealing. It should be noted that the small oxygen pick up in this case is negligible in respect of quasicrystalline phase formation

To obtain further information on the phases mentioned above, we carried out transmission electron microscopic observations. Fig. 5(a,b)  shows the typical selected area diffraction pattern (SADP) of  as-cast $Al_{65}Ga_5Pd_{17}Mn_{13}$ alloy ingot taken from gray region of the microstructure. The typical SADP pattern taken along (a) [010] direction (close to the five fold zone axis of the icosahedral phase) (b) [101] direction confirms that presence of ξ' (orthorhombic structure) phase in the as cast alloy. It is well known that the alloy Al–Pd–Mn of composition $Al_{74}Pd_{22}Mn_4$ has an orthorhombic structure [18]. Some regions corresponding to a B2 cubic phase with a lattice parameter equal to 2.99 Å are also present (the diffraction pattern is not given here). Another type of diffraction pattern has been observed, which corresponds to icosahedral phase along with five fold direction (Fig.5c).

Figure 6 (a-c) shows typical selected area electron diffraction patterns of the $Al_{65}Ga_5Pd_{17}Mn_{13}$ icosahedral quasicrystalline phase , in annealed alloy (at 800°C for 60 h) taken along (a) five-fold axis, (b) three fold axis , (c) two fold axis . The corresponding microstructure is shown in Fig.6 (d). The results presented here suggest the existence of a field of stability for quaternary icosahedral quasicrystals in Al-Ga-Mn-Pd at temperature of nearly 800°C and for compositions around $Al_{65}Ga_5Pd_{17}Mn_{13}$. The quasicrystal is formed when as cast alloy with three type of phases (B2 type, ξ' crystalline and IQC) is annealed at 800°C for 60 h. In annealed specimens only icosahedral quasicrystal has been observed. This suggests that the icosahedral quasicrystal is stable at higher temperatures compared to the others crystalline phase. Upon annealing at 800°C the orthorhombic Al-Ga-Pd-Mn structure (ξ' type phase) as well as B2 phase were converted into icosahedral Al-Ga-Pd-Mn quasicrystal probably through peritectic reaction. It should be mentioned here that the alloy $Al_{65}Ga_5Pd_{17}Mn_{13}$ exhibiting the icosahedral phase has higher Mn concentration than the alloy reported by Fisher *et al.* (1999) [20] i.e. $Al_{67}Ga_4Pd_{21}Mn_8$. It is not clear at present whether this difference in Mn concentration is due different methods of synthesis of the alloy and this aspect requires further investigation

## 4. Conclusion

In the present study we have obtained nearly pure icosahedral phase in Al-Ga-Pd-Mn alloy by annealing the as cast alloy at 800 °C for 60 hours. It has also been found that after annealing the $Al_{65}Ga_5Pd_{17}Mn_{13}$ alloy at 900°C for 60 hours, the icosahedral phase starts decomposing into $Al_3Pd_2$ type phase and icosahedral phase.

## Acknowledgement

The authors would like to thank, Prof. S.Ranganathan, Prof. Ramachandra Rao, Prof.G.V.S.Sastry, Prof. N.K.Mukhopadhyay Prof. J.M.Dudois and Prof. R.K.mandal for many stimulating discussions.  The financial support from CEFIPRA is gratefully acknowledged. One of the authors (T.P.Y) acknowledges to CSIR for a Senior Research Fellowship, during which period the work was completed.

**Figure captions**

**Fig.1** Scanning electron micrograph of (a) as-cast $Al_{65}Ga_5Pd_{17}Mn_{13}$ alloy, showing rod and spheroid like features which are marked A and B respectively  (b) The microstructure  obtained after annealing at 800 °C for 60 h,

Fig.2 DTA scan of as cast $Al_{65}Ga_5Pd_{17}Mn_{13}$ alloy in vacuum. The endothermic peak at ~ 760°C may be noticed.

Fig.3 X-ray diffraction pattern obtained from (a) as cast $Al_{65}Ga_5Pd_{17}Mn_{13}$ alloy (b) annealed at 700° C for 60, (c)  annealed at 800 °C for 60h.

Fig.4  Energy–dispersive spectrum of the as cast $Al_{65}Ga_5Pd_{17}Mn_{13}$ alloy.

Fig.5, Selected area diffraction patterns of as cast $Al_{65}Ga_5Pd_{17}Mn_{13}$ alloy (a-c) and the corresponding microstructure (d).

Fig.6 The SADPs of annealed version of $Al_{65}Ga_5Pd_{17}Mn_{13}$ alloy, (a) five-fold (b) three fold (c) two fold and corresponding bright field microstructure respectively.

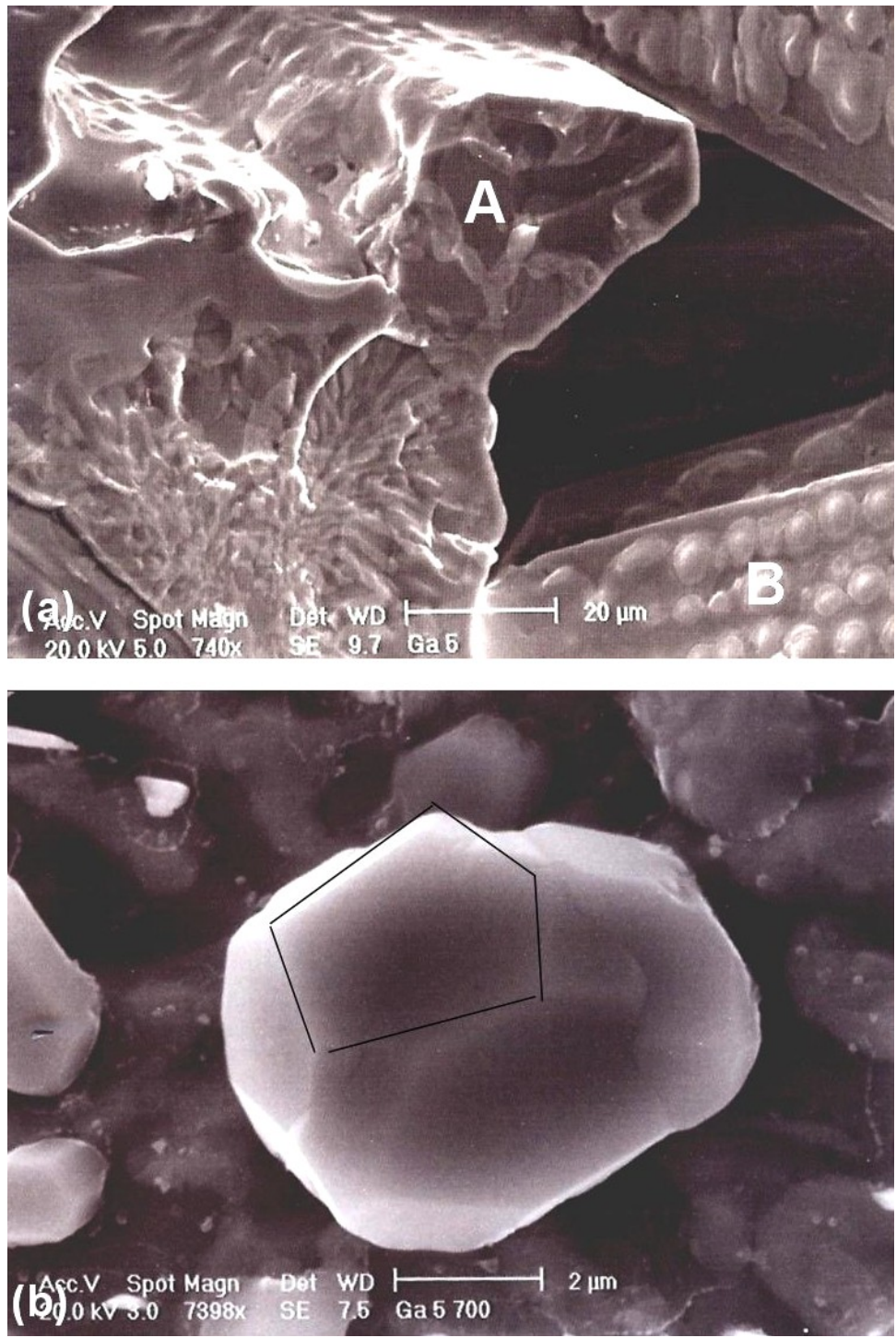

Fig.1

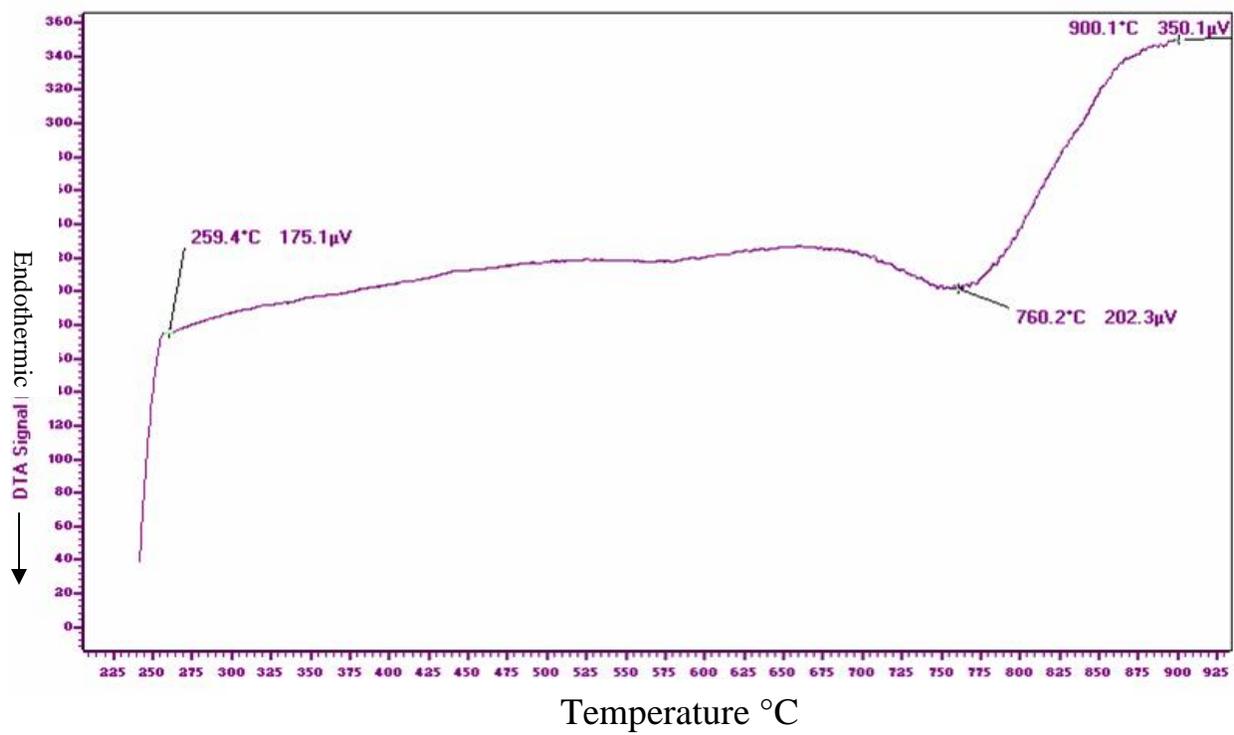

Fig.2

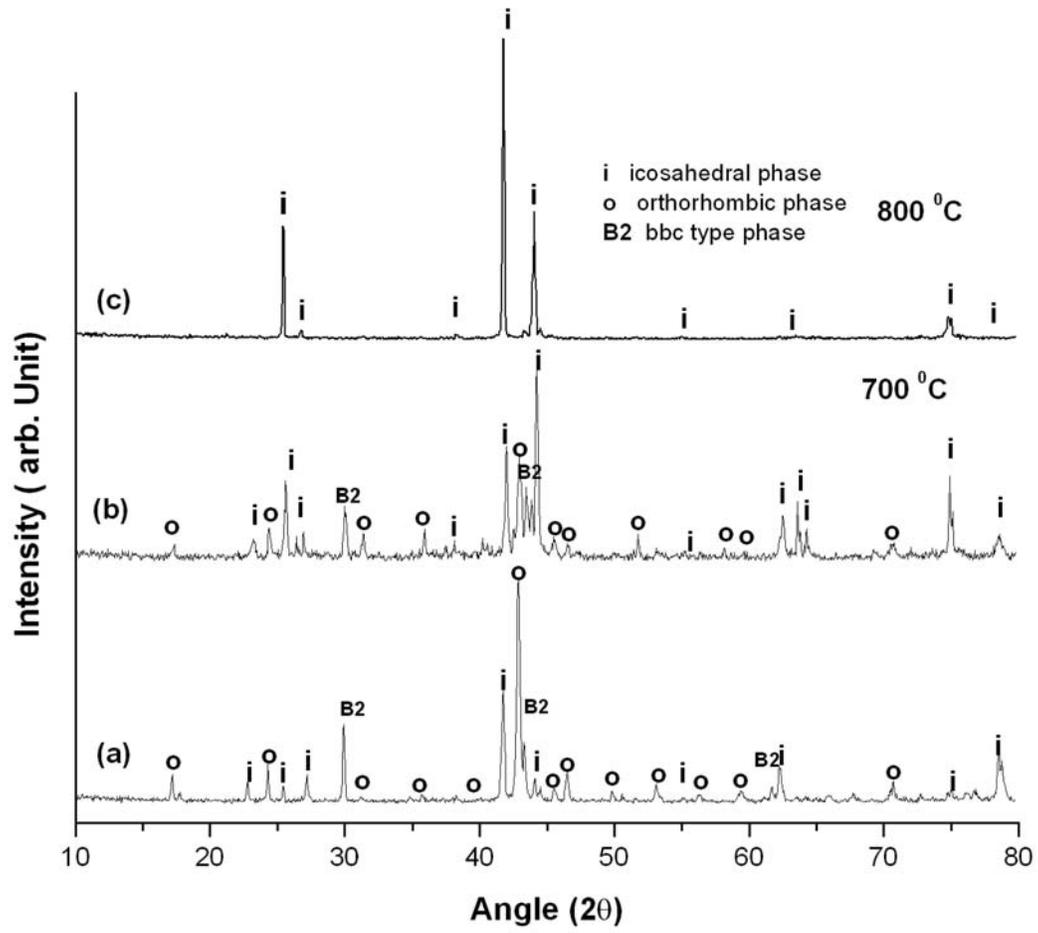

Fig.3

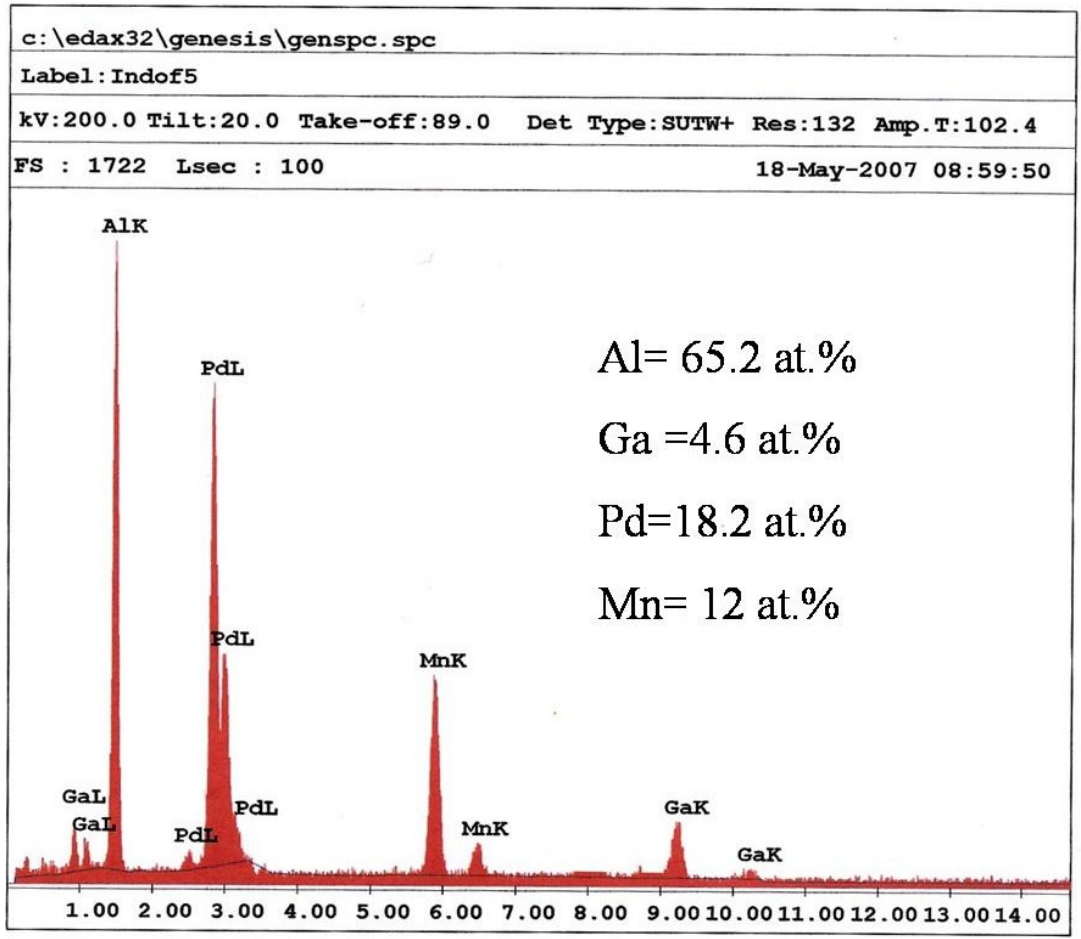

Fig.4

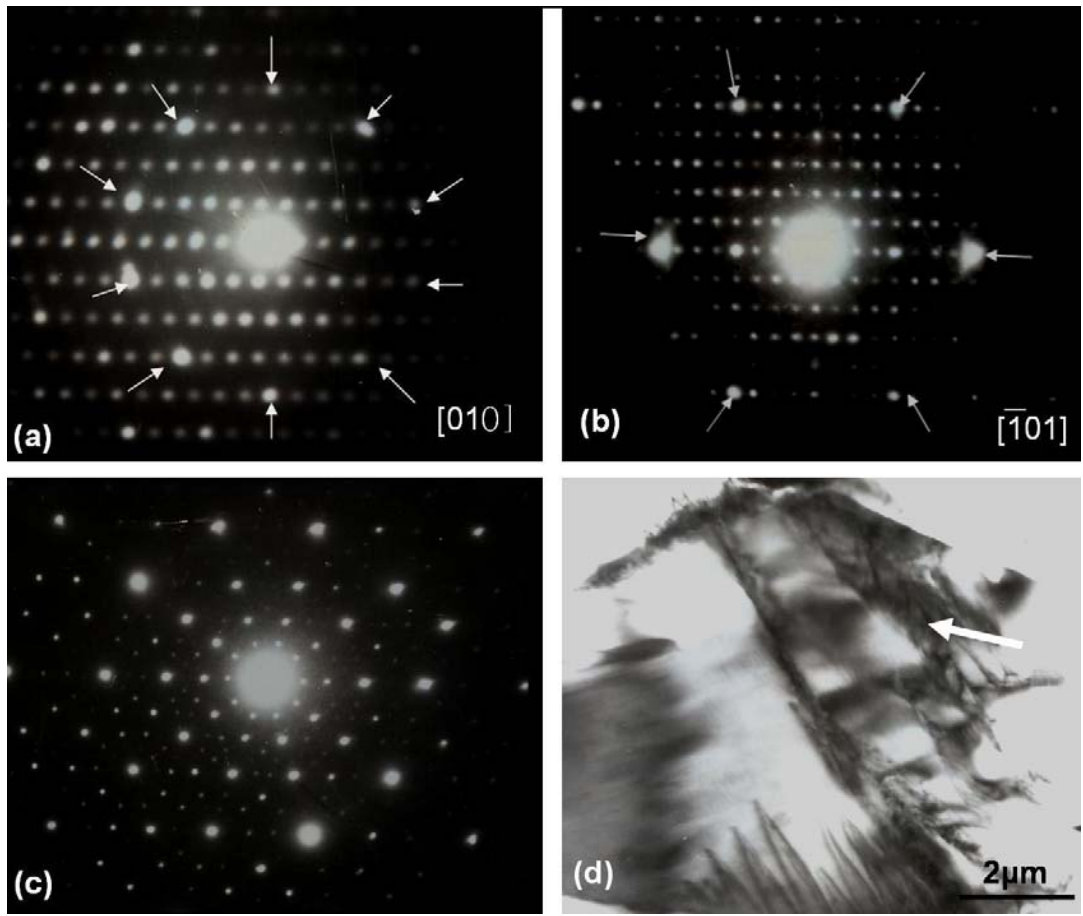

(a) [010]

(b) [$\bar{1}$01]

(c)

(d) 2μm

Fig.5

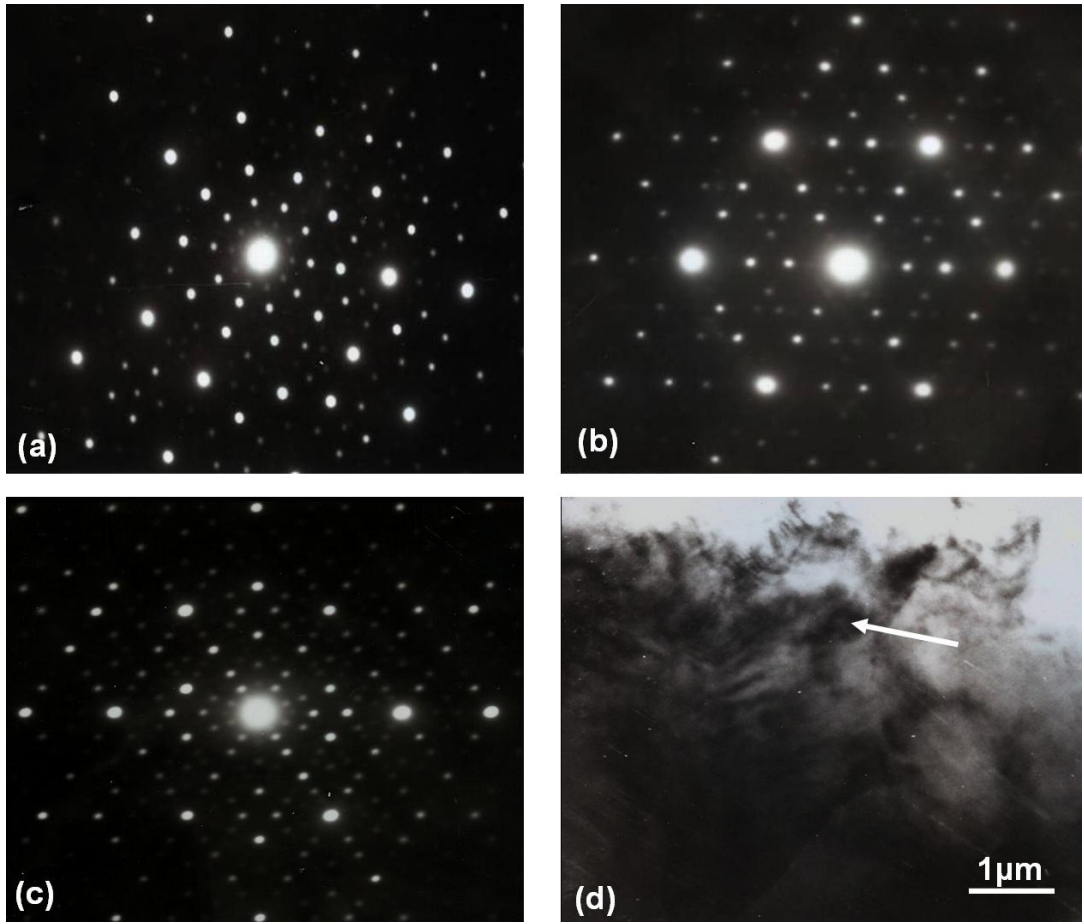

Fig.6